\documentclass[letterpaper,english,prl,twocolumn,superscriptaddress,showpacs,preprintnumbers]{revtex4}

\usepackage{verbatim}
\usepackage{graphicx}

\makeatletter

\newcommand{\lyxdot}{.}

\usepackage{dcolumn}
\usepackage{bm}

\newcommand{\be}{\begin{equation}}
\newcommand{\ee}{\end{equation}}

\usepackage{babel}
\makeatother
\begin{document}

\title{Quantum critical scaling of the geometric tensors}

\author{Lorenzo Campos Venuti}

\affiliation{Institute for Scientific Interchange, Villa Gualino, Viale Settimio
Severo 65, I-10133 Torino, Italy }

\author{Paolo Zanardi}

\affiliation{Department of Physics and Astronomy, University of Southern California
Los Angeles, CA 90089-0484 (USA) }

\affiliation{Institute for Scientific Interchange, Villa Gualino, Viale Settimio
Severo 65, I-10133 Torino, Italy }

\date{\today}

\begin{abstract}
Berry phases and the quantum-information theoretic notion of {\em
fidelity} have been recently used to analyze quantum phase transitions
from a geometrical perspective. In this paper we unify these two approaches
showing that the underlying mechanism is the critical singular behavior
of a complex tensor over the Hamiltonian parameter space. This is
achieved by performing a scaling analysis of this quantum geometric
tensor in the vicinity of the critical points. In this way most of
the previous results are understood on general grounds and new ones
are found. We show that criticality is not a suffcient condition to
ensure superextensive divergence of the geometric tensor, and state
the conditions under which this is possible. The validity of this
analisys is further checked by exact diagonalisation of the spin-$1/2$
$XXZ$ Heisenberg chain. 
\end{abstract}


\maketitle
{\em Introduction.--} Phase transitions at zero-temperatures are
dramatic changes in the ground-state (GS) properties of a system driven
by quantum fluctuations. This phenomenon, known as quantum phase transition
(QPT), is due to the interplay between different orderings associated
to competing terms in the system's Hamiltonian \cite{sachdev}. %
{} Traditionally such a problem is addressed by resorting to notions
like order-parameter and symmetry breaking i.e., the Landau-Ginzburg
paradigm \cite{gold}. In the last few years a big effort has been
devoted to the analysis of QPTs from the perspective of Quantum Information
\cite{qis} the main tool being the study of different entanglement
measures \cite{qpt_qis}.

More recently an approach to QPTs based on the quantum-information
concept of {\em fidelity} has been put forward \cite{za-pa}, \cite{zhou}.
The strategy there is differential-geometric and information-theoretic
in nature: GSs associated to infinitesimally close parameters are
compared i.e., their overlap evaluated. The intuition behind is extremely
simple: at QPTs even the slightest move results in a major difference
in some of the system's observables, in turn this latter has to show
up in the degree of orthogonality i.e., fidelity, between the corresponding
GSs. Systems of quasi-free fermions have been analyzed \cite{za-co-gio},\cite{co-gio-za}
as well as QPTs in matrix-product states \cite{co-ion-za}. Finite-temperature
extensions have been also considered showing the robustness of the
approach against mixing with low excited states \cite{zhong-guo}.
Remarkably the fidelity analysis have been successfully carried over
for the superfluid-insulator transition of the Hubbard model \cite{pier};
this suggests that this framework, besides its conceptual appeal,
can have some practical relevance even for fully-interacting systems
where a simple description is not possible.

In Ref \cite{DG-qpt} it has been shown that the fidelity approach
can be better understood in terms of a Riemannian metric tensor $g$
defined over ${\cal M}$. Loosely speaking the singularities developed
by $g$ in the thermodynamic limit correspond to QPTs \cite{DG-qpt}.
Even for finite-size systems, critical points have a markedly distinct
(finite-size scaling) behavior from the regular ones; in all the example
studied so far this difference amounts to an enhanced orthogonalization
ratio as a function of the system's size at the QPTs. Another intriguing
relation between QPTs and geometrical objects i.e.~Berry-phases,
was suggested in Refs \cite{BP-qpt0} and \cite{BP-qpt1}. There it
was argued that loops in the parameter space, encircling a critical
line give rise to a non zero GS Berry phase even for an arbitrary
small loop size. This fact indicates that at the critical points the
curvature of the Berry connection should display some sort of singularity\cite{BP-qpt1}.

In this paper we shall show that these two approaches share the same
origin and can be therefore unified. We will perform a scaling analysis
that allows one to understand, from a single perspective, most of
the results obtained so far in the fidelity approach and to investigate
somewhat unexpected new ones. 

{\em Geometric tensors.--} We now lay down the formal setting.
For each element $\lambda$ of the parameter manifold ${\cal M}$
there is an associated quantum Hamiltonian $H(\lambda)=\sum_{n=0}^{{\rm {dim}{\cal H}}}E_{n}(\lambda)|\Psi_{n}(\lambda)\rangle\langle\Psi_{n}(\lambda)|,\,(E_{n+1}\ge E_{n})$
acting over a finite-dimensional state-space ${\cal H}$; the mapping
$\lambda\rightarrow H(\lambda)$ is assumed to be smooth. If $|\Psi_{0}(\lambda)\rangle$
denotes the unique ground state (GS) of $H(\lambda)$ then one has
the mapping $\Psi_{0}\colon{\cal M}\rightarrow{\cal H}\colon\lambda\rightarrow|\Psi_{0}(\lambda)\rangle.$
More properly we will consider this map as a function valued in the
projective space $P{\cal H}$ of rays. If the spectral gap of $H(\lambda)$
above the GS is bounded away from zero over ${\cal M}$ then $\Psi_{0}$
is smooth \cite{bathia}. The projective Hilbert space is the base
manifold of a $U(1)$ fiber bundle \cite{Nak} and it is equipped
with a complex metric given by $g(u,v)=\langle u,(1-|\Psi\rangle\langle\Psi|)v\rangle$
($u$ and $v$ denote tangent vectors in $|\Psi\rangle\langle\Psi|$).
Pulling this metric back to ${\cal M}$ by $\Psi_{0}$ i.e., evaluating
it on vectors of the form $d/dt\Psi_{0}(\lambda(t))$ one obtains
the complex hermitean tensor \begin{equation}
Q_{\mu\nu}:=\langle\partial_{\mu}\Psi_{0}|\partial_{\nu}\Psi_{0}\rangle-\langle\partial_{\mu}\Psi_{0}|\Psi_{0}\rangle\langle\Psi_{0}|\partial_{\nu}\Psi_{0}\rangle\label{Q_munu}\end{equation}
 Here the indices $\mu$ and $\nu$ are labeling the coordinates of
${\cal M}$ i.e., $\mu,\nu=1,\ldots,{\rm {dim}\,{\cal M}.}$ This
quantity is the {\em quantum geometric tensor} (QGT) \cite{pro},
both its real and imaginary parts have a relevant physical meaning.

The real part $g_{\mu\nu}:=\Re Q_{\mu\nu}$ is a {\em Riemannian}
(real) metric tensor over ${\cal M}$ which defines the line element
as $ds^{2}=\sum_{\mu\nu}g_{\mu\nu}d\lambda_{\mu}d\lambda_{\nu}.$
This metric has been shown to provide the leading term in the expansion
of the fidelity between two GSs associated to slightly different Hamiltonians
\cite{DG-qpt}. More precisely if ${\cal F}(\lambda,\lambda^{\prime}):=|\langle\Psi_{0}(\lambda),\Psi_{0}(\lambda^{\prime})\rangle|$
is the fidelity then 
${\cal F}(\lambda,\lambda+\delta\lambda)\approx1-{\delta\lambda^{2}}/{2}\, g(d\Psi_{0},d\Psi_{0})=1-ds^{2}/{2}$
(i.e., $g_{\mu\nu}$ is the Hessian matrix of ${\cal F}(\lambda,\lambda^{\prime})$
as a function of $\lambda^{\prime}$ in $\lambda=\lambda^{\prime}$).
 The meaning of this distance function between parameters should be
obvious: it is the Hilbert-space one between the corresponding GSs.
This latter quantifies the operational distinguishability of two states
\cite{Woo}; therefore even the induced metric $g$ conveys a definite
information-theoretic meaning \cite{DG-qpt}. The fact that at the
QPTs $g$ exhibits singularities is consistent with the intuition
that at critical points one has a major change in the GS structure
i.e., it becomes \char`\"{}more different\char`\"{}, and makes it
quantitative. 
Now we consider the imaginary part of (\ref{Q_munu}) $F_{\mu\nu}:=\Im Q_{\mu\nu}.$
Since the terms $\langle\Psi|\partial_{\nu}\Psi\rangle$ are -from
normalization- purely imaginary, one finds $\Im Q_{\mu\nu}=\Im\langle\partial_{\mu}\Psi|\partial_{\nu}\Psi\rangle=\langle\partial_{\mu}\Psi|\partial_{\nu}\Psi\rangle-\langle\partial_{\nu}\Psi|\partial_{\mu}\Psi\rangle=\partial_{\mu}A_{\nu}-\partial_{\nu}A_{\mu},$
where $A_{\mu}:=\langle\Psi|\partial_{\nu}\Psi\rangle$ is, for $|\Psi\rangle=|\Psi_{0}(\lambda)\rangle,$
the Berry adiabatic connection \cite{BP}. From this one sees that
$\Im Q_{\mu\nu}$ is nothing but the curvature 2-form, responsible
for the appearance of the Berry geometrical phase \cite{BP}. Of course
for systems with real GS $F$ is zero and the QGT coincides with its
real part $g$. 

The QGT (\ref{Q_munu}) can be cast in a way useful for later derivations
as well as to decrypt its physical meaning. By inserting in Eq. (\ref{Q_munu})
the spectral resolution $\openone=\sum_{n=0}^{{\rm {dim}{\cal H}}}|\Psi_{n}(\lambda)\rangle\langle\Psi_{n}(\lambda)|$
and differentiating the eigenvalue equation $H(\lambda)|\Psi_{0}(\lambda)\rangle=E_{0}(\lambda)|\Psi_{0}(\lambda)\rangle,$
one finds the identity \begin{equation}
Q_{\mu\nu}=\sum_{n\neq0}\frac{\langle\Psi_{0}(\lambda)|\partial_{\mu}H|\Psi_{n}(\lambda)\rangle\langle\Psi_{n}(\lambda)|\partial_{\nu}H|\Psi_{0}(\lambda)\rangle}{[E_{n}(\lambda)-E_{0}(\lambda)]^{2}}\ .\label{Q-pert}\end{equation}
 This expression clearly suggests that at the critical points, where
one of the $\varepsilon_{n}(\lambda_{c})=E_{n}(\lambda_{c})-E_{0}(\lambda_{c})\ge0$
vanishes in the thermodynamic limit, the QGT might show a singular
behavior. This heuristic argument is the same one proposed in (\cite{DG-qpt})
for the Riemannian tensor $g_{\mu\nu}$ and for the Berry Curvature
$F_{\mu\nu}$ in \cite{BP-qpt1}. One of the aims of this paper is
to establish this argument on more firm grounds. A similar scaling
analysis in connection with local measures of entanglement at QPTs
has been presented in \cite{lcv-LME}.

We would like first to demonstrate an inequality useful to establish
a connection between the tensor $g$ and QPTs. We consider a system
with size $L^{d}$ ($d$ is the spatial dimension). Since $Q(\lambda)$
is an hermitean non-negative matrix one has $|Q_{\mu\nu}|\le\| Q\|_{\infty}=\langle\phi|Q|\phi\rangle$
where $|\phi\rangle=(\phi_{\mu})_{\mu=1}^{{\rm {dim}}{\cal M}}$ denotes
the eigenvector of $Q$ corresponding to the largest eigenvalue. We
set $\delta H=\sum_{\mu}(\partial_{\mu}H)\phi_{\mu},$ then from Eq.
(\ref{Q-pert}) and the above inequality \begin{eqnarray}
|Q_{\mu\nu}| & \le & \sum_{n>0}\varepsilon_{n}^{-2}|\langle\Psi_{0}|\delta H|\Psi_{n}\rangle|^{2}\le\varepsilon_{1}^{-2}\sum_{n>0}|\langle\Psi_{0}|\delta H|\Psi_{n}\rangle|^{2}\nonumber \\
 & = & \varepsilon_{1}^{-2}(\langle\delta H\delta H^{\dagger}\rangle-|\langle\delta H\rangle|^{2}),\label{ineq}\end{eqnarray}
where the angular brackets denote the average over $|\Psi_{0}(\lambda)\rangle$.
Now we assume that the operator $\delta H$ is a local one i.e., $\delta H=\sum_{j}\delta V_{j}$;
then the last term in Eq. (\ref{ineq}) reads $\sum_{i,j}(\langle\delta V_{i}\delta V_{j}^{\dagger}\rangle-\langle\delta V_{i}\rangle\langle\delta V_{j}^{\dagger}\rangle)$
If the GS is translationally invariant this last quantity can be written
as $L^{d}\sum_{r}K(r):=L^{d}K$ where $K(r):=\langle\delta V_{i}\delta V_{i+r}^{\dagger}\rangle-\langle\delta V_{i}\rangle\langle\delta V_{i+r}^{\dagger}\rangle$
independent on $i.$ For gapped systems i.e., $\varepsilon_{1}(\infty):=\lim_{L\to\infty}\varepsilon_{1}(L)>0$
the correlation function $G(r)$ is rapidly decaying \cite{hastings}
and therefore $K$ is finite and independent on the system size. Using
(\ref{ineq}) it follows that for these non-critical systems $|Q_{\mu\nu}|$
{\em cannot grow, as a function of $L,$ more than extensively}.
Indeed one one has that 
$\lim_{L\to\infty}|Q_{\mu\nu}|/L^{d}\le K\varepsilon_{1}^{-2}(\infty)<\infty.$
Conversely if $\lim_{L\to\infty}|Q_{\mu\nu}|/L^{d}=\infty$ i.e.,
$|Q_{\mu\nu}|$ grows super-extensively, then either $\varepsilon_{1}(L)\rightarrow0$
or $K$ cannot be finite. In both cases the system has to be gapless
\cite{complete}. Summarizing: {\em a super-extensive behavior of
any of the components of $Q$ for systems with local interaction implies
a vanishing gap in the thermodynamic limit. }

This sort of behavior has been observed in all the systems analyzed
in Refs \cite{za-pa}--\cite{co-ion-za} and does amount to the critical
fidelity drop at the QPTs. As we will show in the next section, in
general the converse result i.e., QPT$\rightarrow$ super-extensive
growth of $ds^{2}(L)$ does not hold true: {\em $Q_{\mu\nu}/L^{d}$
can be finite in the thermodynamic limit even for gapless systems.}
In order to demonstrate this fact we now move to a representation
of $Q_{\mu\nu}$ in terms of suitable correlation functions. This
key move is an extension of the results of You {\em et al.} \cite{gu},
for the so-called fidelity susceptibility, to the whole QGT.

{\em Correlation Function representation.--} Let us consider the
following imaginary time (connected) correlation functions \begin{equation}
G_{\mu\nu}(\tau)=\theta(\tau)\left(\langle\partial_{\mu}H(\tau)\partial_{\nu}H(0)\rangle-\langle\partial_{\mu}H(0)\rangle\langle\partial_{\nu}H(0)\rangle\right),\label{corr}\end{equation}
 where $X(\tau):=e^{\tau H}Xe^{-\tau H}.$ Using again the spectral
resolution of the identity associated to $H(\lambda)$ one finds $G_{\mu\nu}(\tau)=\theta(\tau)\sum_{n>0}e^{-\varepsilon_{n}(\lambda)\tau}X_{n\mu}X_{n\nu}^{*},$
where $X_{n\mu}:=\langle\Psi_{0}(\lambda)|\partial_{\mu}H|\Psi_{n}(\lambda)\rangle.$
Notice that if $H(\lambda)=H_{0}+\lambda V$ then $G(\tau)$ is nothing
but the dynamic response function associated to the \char`\"{}perturbation\char`\"{}
$V.$ We now move to the frequency domain 
$\tilde{G}_{\mu\nu}(\omega)=\int_{-\infty}^{+\infty}d\tau e^{-i\omega\tau}G_{\mu\nu}(\tau)=\sum_{n>0}X_{n\mu}X_{n\nu}^{*}(i\omega+\varepsilon_{n})^{-1}$
By comparing this equation with (\ref{Q-pert}) it is immediate to
see that \begin{equation}
Q_{\mu\nu}=-i\frac{d}{d\omega}\tilde{G}_{\mu\nu}(\omega)|_{\omega=0}=\int_{-\infty}^{+\infty}d\tau\tau G_{\mu\nu}(\tau).\label{corr-rep}\end{equation}
 This equation is an integral representation of the QGT in terms of
the (imaginary time) correlation functions of the operators $\partial_{\mu}H.$
 Eq (\ref{corr}) is remarkable in that it connect the tensors $g_{\mu\nu}$
and $F_{\mu\nu}$ directly (and non perturbatively) to the dynamical
response of the system to the interactions $\partial_{\mu}H$'s. In
this way geometrical and information-theoretic objects $F$ and $g$
are expressed in terms of standard quantities in response theory and
their physics content is so further clarified. Eqs. (\ref{corr})
and (\ref{corr-rep}) provide the starting point for our scaling analysis

{\em Scaling behavior.--} First we assume that the operators $\partial_{\mu}H$
are local ones i.e., $\partial_{\mu}H=\sum_{x}V_{\mu}(x).$We also
rescale the QGT (\ref{Q_munu}) by the system size $Q_{\mu\nu}\rightarrow q_{\mu\nu}=L^{-d}Q_{\mu\nu}$
in order to obtain well-defined quantities in the thermodynamic limit.
Now we consider the scaling transformations $x\rightarrow\alpha x,\,\tau\rightarrow\alpha^{\zeta}\tau,(\alpha\in{\bf {R}}^{+}).$
Assuming that, in the vicinity of the critical point $\lambda_{c}$,
the operators $V_{\mu}$ have well-defined scaling dimensions \cite{scal-rel}
one has $V_{\mu}\rightarrow\alpha^{-\Delta_{\mu}}V_{\mu};$ these
relations along with Eqs. (\ref{corr}) and (\ref{corr-rep}) imply
\begin{equation}
q_{\mu\nu}\rightarrow\alpha^{-\Delta_{\mu\nu}^{Q}}q_{\mu\nu};\quad\Delta_{\mu\nu}^{Q}:=\Delta_{\mu}+\Delta_{\nu}-2\zeta-d\,.\label{Q-scal}\end{equation}
 For simplicity we assume now that there is only one driving parameter
$\lambda$ and drop the indices $\mu$ and $\nu$ in $\Delta_{\mu\nu}^{Q}.$

If $\xi$ is the correlation length one has $\xi=|\lambda-\lambda_{c}|^{-\nu}$
(here $\nu$ is the correlation length critical exponent) and, if
$\Delta_{\lambda}$ is the scaling dimension of the driving parameter
$\lambda,$ $\nu=\Delta_{\lambda}^{-1}.$ Putting all this together
and following standard arguments in scaling analysis one obtains that
(in the off-critical region, $L\gg\xi$) the singular part of the
--intensive-- QGT behaves as \begin{equation}
q_{\mu\nu}(\lambda\approx\lambda_{c})\sim|\lambda-\lambda_{c}|^{\Delta_{Q}/\Delta_{\lambda}}.\label{Q-lambda}\end{equation}
 Instead at the critical point i.e.~$\xi=\infty$, where the only
length scale is provided by the system size itself one gets \begin{equation}
q_{\mu\nu}(\lambda=\lambda_{c})\sim L^{-\Delta_{Q}}.\label{Q-L}\end{equation}
 Equations (\ref{Q-lambda}) and (\ref{Q-L}) represent the main result
of this paper. From Eq. (\ref{Q-lambda}) one sees that close to the
critical point the QGT $q_{\mu\nu}$ is diverging for $\Delta_{Q}/\Delta_{\lambda}<0$;
on the other hand when one is sitting exactly at the critical point
one finds that, besides an extensive contribution coming from the
regular part, the singular part contributes to $Q_{\mu\nu}$ in a
manner which is: i) super-extensive if $\Delta_{Q}<0$ ii) extensive
if $\Delta_{Q}=0$ and iii) sub-extensive for $\Delta_{Q}>0.$ Hence
we observe that {\em $q_{\mu\nu}$ can be finite at the critical
point, even in a gapless system, provided $\Delta_{Q}>0$}. An explicit
example of this phenomenon will be discussed in the sequel; before
doing that we show that this analysis allows one to understand in
a unified manner the results found for quasi-free fermionic models
in \cite{za-co-gio},\cite{co-gio-za}. In quasi-free fermionic models
the most relevant operator admissible has scaling dimension equal
to one, therefore from (\ref{Q-lambda}) and (\ref{Q-L}) one finds,
close to $\lambda_{c}$ , $Q_{\mu\nu}=g_{\mu\nu}\sim L|\lambda-\lambda_{c}|^{-1}$
and $Q_{\mu\nu}\sim L^{2}$ at the QPT. Notice that if $\Delta Q<-1$
one expects super-quadratic behavior. 

{\em $XXZ$ chain}.-- We provide now a further test for the ideas
presented in this paper: the $S=1/2$ $XXZ$ Heisenberg chain. The
model is defined by $H=J\sum_{i}\left[S_{i}^{x}S_{i+1}^{x}+S_{i}^{y}S_{i+1}^{y}+\lambda S_{i}^{z}S_{i+1}^{z}\right]$.
It is well known (see e.g.~\cite{gogolin}) that in the regime $\lambda\in(-1,1)$
the model is in the universality class of a $c=1$ conformal field
theory ($d=\zeta=1$) displaying gapless excitations and power low
correlations. For $\lambda>1$ the model enters a phase with Ising-like
antiferromagnetic order and a non-zero gap. Finally, the isotropic
point $\lambda=1$ is a Berezinskii-Kosterlitz-Thouless transition
point. The low energy effective continuum theory is given by the sine-Gordon
model: 
$H=\int d^{2}x\left\{ \frac{u}{2}\left[\Pi^{2}+\left(\partial_{x}\Phi\right)^{2}\right]-\frac{v\lambda}{\left(a\pi\right)^{2}}\cos\left(\sqrt{16\pi K}\Phi\right)\right\} $
here $v$ is the bare Fermi velocity, $u$ the renormalized one, $K$
is related to the compactification radius of the field $\Phi$ and
$a$ is the lattice spacing (we use the notations of \cite{gogolin}).

We now analyze the behavior of the fidelity when the anisotropy parameter
$\lambda$ is varied. Correspondingly we are interested in the operator
$V\left(x\right)=S_{x}^{z}S_{x+1}^{z}$ and its scaling exponents.
In the continuum limit the operator $V\left(x\right)$ contributes
with a marginal operator -- $\left(\partial_{x}\Phi\right)^{2}$ --
with scaling exponent $\Delta_{V}=2$, plus a term which is precisely
the cosine in the sine-Gordon Hamiltonian. The cosine term is irrelevant
for $\left|\lambda\right|<1$, marginal at $\lambda=1$ and relevant
for $\lambda>1$ where it is responsible for the opening of a mass
gap. Its scaling dimension is precisely equal to $4K$. The parameter
$K$ can be fixed by the long distance asymptotic of the correlation
functions obtained by Bethe Ansatz, and one gets, for $\lambda\in(-1,1]$
$K={\pi/2}{\left(\pi-\arccos\left(\lambda\right)\right)^{-1}}.$

When one considers the finite size scaling of $q$ in the gapless
region $\left|\lambda\right|<1$, the leading contribution is dictated
by the most relevant component of $V\left(x\right)$ which in this
case is the marginal one i.e.~$\Delta_{V}=2$. Correspondingly, using
equation (\ref{Q-L}) we obtain $\Delta_{Q}=1$, so that, in the whole
region $\left|\lambda\right|<1$, the finite size dependence of the
(rescaled) QGT tensor is $q=A_{1}+A_{2}L^{-1}$, where $A_{1,2}$
are constants which depend only on $\lambda$. Note that the term
$A_{1}$ is the contribution coming from the regular part. This kind
of scaling has to be contrasted with the one observed in quasi-free
fermionic systems, where one has $q={A'}_{1}+{A'}_{2}L$. We would
like to stress again that in general {\em one expects a super-extensive
behavior of $Q,$ and a corresponding fidelity drop when $d+2\zeta-2\Delta_{V}>0$
i.e., when the operator associated to the varying parameter is sufficiently
relevant.} When this condition is not fulfilled the rescaled QGT
tensor does not diverge in the thermodynamic limit at critical points.
Nevertheless a proper finite size scaling analysis allows one to identify
the critical region. To check this latter feature as well as the predicted
scaling behavior we have performed exact Lanczos diagonalizations.
The agreement between numerical data and the theoretical prediction
is shown in Fig (\ref{fig:fit}). %
{}

\begin{center}%
\begin{figure}
\begin{centering}\includegraphics[width=7cm,keepaspectratio]{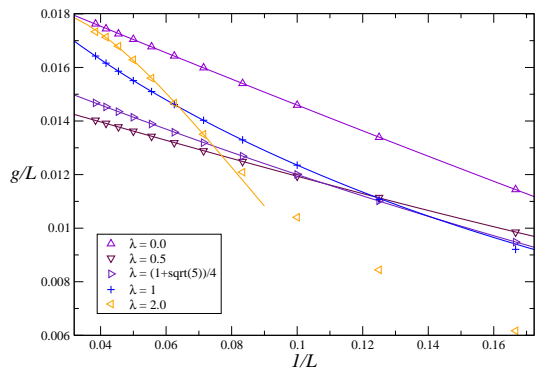}\par\end{centering}

\caption{Scaling behavior of metric $g$. The points are the data obtained
via Lanczos diagonalization of periodic chains of length $L$ up to
26, using equation ${\cal F}(\lambda,\lambda+\delta\lambda)\approx1-g\delta\lambda^{2}/{2}$
with $\delta\lambda=1.0\times10^{-3}$. The points at $\left|\lambda\right|<1$
are well fitted (solid lines) with $g/L=A_{1}+A_{2}L^{-1}+A_{3}L^{3-2\Delta_{V}^{\left(2\right)}}$.
The contribution $A_{3}$ comes from the irrelevant operator with
scaling dimension $\Delta_{V}^{\left(2\right)}=4K$. As this operator
becomes rapidly irrelevant for $\left|\lambda\right|<1$ its contribution
can be hardly observed. At $\lambda=1$ a better a fit is obtained
with logarithmic corrections as expected at the isotropic point. In
the massive regime we expect that the thermodynamic limit is approached
exponentially fast. We obtain a good agreement by fitting our data
with the phenomenological formula $g/L=A_{1}+A_{2}e^{-L/\xi}L^{-1/2}$
where $\xi$ is the correlation length as given by Bethe Ansatz \cite{baxter}.
As $\xi\left(\lambda=2\right)=8.35\ldots$ we used only points with
$L\ge14$.\label{fig:fit}}
\end{figure}
\par\end{center}

{\em Conclusions.--} The mapping between a quantum Hamiltonian
and the corresponding ground state endows the parameter manifold with
a complex tensor $Q.$ The real part of $Q$ is a Riemannian metric
$g$ while the imaginary part is the curvature form giving rise to
a Berry phase. The metric tensor $g$ is closely related to the quantum
fidelity between different ground states; in the thermodynamical limit
it has been shown to be singular at the critical point for several
models featuring quantum phase transitions e.g., quasi-free fermionic
models. The same kind of singularity have been argued to exist for
the form $F$ and the associated Berry phases as well. In this paper
we demonstrated that i) the components $Q_{\mu\nu}$ of $Q$ have
an integral representation in terms of response functions ii) a super-extensive
behavior of any of the $Q_{\mu\nu}$'s implies, for local models,
gaplessness iii) the singular part of the $Q_{\mu\nu}$ fulfills scaling
relations which explicitly connect their singular behavior with the
universality class of the transition i.e., critical exponents. In
particular these relations show that gaplessness is a necessary but
not sufficient condition for a super-extensive scaling of the metric
tensor, i.e.~enhanced orthogonalization rate. 
The theoretical analysis has been further supported by a numerical
study of the finite-size scaling of the fidelity for the $XXZ$ spin
$1/2$ chain. The main message of this paper is that apparently unrelated
results can be understood in unified fashion by unveiling the underlying
common differential-geometric structure and analyzing its quantum-critical
behavior.

{\em Acknowledgments} LCV wishes to thank H. Nishimori for providing
the Lanczos code \texttt{TITPACK.} We thank C.~Degli Esposti Boschi,
M.~Roncaglia, M. Cozzini and A. Hamma for valuable discussions. 
\acknowledgments


\end{document}